\documentclass{eas}
\usepackage{graphicx}
\usepackage{astron}
\usepackage{amsmath}
\usepackage{amssymb}
\begin{document}

\title{How the mass-loss rates of red-supergiants determine the fate of massive stars ?} 
\runningtitle{Mass loss from RSGs and later evolution}
\author{Cyril Georgy}\address{CRAL, ENS-Lyon, 46  All\'ee d'Italie, 69364 Lyon Cedex 07, France}
\author{Sylvia Ekstr\"om}\address{Geneva Observatory, Geneva University, Chemin des Maillettes 51, 1290 Versoix, Switzerland}
\author{Hideyuki Saio}\address{Astronomical Institute, Graduate School of Science, Tohoku University, Sendai, Japan}
\author{Georges Meynet}\sameaddress{2}
\author{Jose Groh}\sameaddress{2}
\author{Anah\'i Granada}\sameaddress{2}
\begin{abstract}
Mass-loss rates are one of the most relevant parameters determining the evolution of massive stars. In particular, the rate at which the star loses mass during the red-supergiant (RSG) phase is the least constrained by the observations or theory. In this paper, we show how the mass loss during the RSG phase affects the later evolution of the star, as well as the final type of supernova towards which it leads. We also discuss some possibilities to discriminate between blue stars that went through a RSG phase and those which remained in the blue part of the Hertzsprung-Russell diagram.
\end{abstract}
\maketitle
\section{Introduction}

Since more than ten years, it has become obvious that rotation, mass loss, and their interplay have a deep impact on the evolution of stars (see for example the review by Maeder \& Meynet \cite*{Maeder2012a}). Both of them contribute to the chemical enrichment of the stellar surface, and the stellar winds then carry away these new chemical elements in the circum-stellar medium (CSM), contributing to the chemical enrichment of the Universe. The exact stellar evolution is however crucially dependent on several not-well-known physical processes and parametrisations, such as the way rotation is implemented in stellar evolution codes, the treatment of magnetic fields, or the mass-loss rates during the various phases of stellar evolution.

The mass-loss rates during the red-supergiant (RSG) phase are still uncertain. In stellar-evolution codes, the mass-loss rates used during this phase are usually those determined by de Jager et al. \cite*{deJager1988a}. Even if these rates have been confirmed by a recent study \cite{Mauron2011a}, some other determinations suggest that in some cases, they can be much higher \cite{vanLoon2005a}. Observations also seem to show that RSGs can lose mass dramatically in some kind of bursts \cite[see also the paper by Humphreys in these proceedings]{Humphreys2005a,Smith2009a,Moriya2011a}. It is unclear what is the total amount of mass lost in the RSG phase, and how an average mass loss rate should be computed from the possibly discrete mass loss episodes.

In this paper, we discuss how the mass loss during the RSG phase impacts the subsequent evolution of the star, and how recent stellar evolution computations fit (or do not fit) the observations, in the framework of single massive star evolution.

\section{Red supergiant, and then?}

\subsection{How rotation and mass loss govern the duration of the RSG phase}

It has been known for long that a RSG star can evolve bluewards, if the ratio $q = \frac{M_\text{core}}{M_\text{tot}}$ of the mass of its He core to the total mass becomes larger than a given value \cite{Giannone1967a}, typically around $0.7$. In the framework of the evolution of a single star, rotation and mass loss affect this ratio as follows:
\begin{itemize}
\item Rotation, particularly during the main sequence (MS), increases the size of the He core, due to the supply of fresh fuel in the core by rotational mixing \cite{Meynet2000a}.
\item Mass loss during the RSG phase contributes obviously to the decrease of the total mass of the star.
\end{itemize}
Both effects contribute to the increase of $q$ during the RSG phase, and thus, favours an evolution towards higher $T_\text{eff}$ after the RSG phase.

\subsection{New models with the Geneva stellar evolution code}

Recently, we computed a new grid of stellar models at solar metallicity \cite{Ekstrom2012a}, with the Geneva stellar evolution code \cite{Eggenberger2008a}. The free parameters of the code (mixing length, overshooting, rotation parametrisation) were calibrated in order to reproduce various observations (width of the MS, red giant branch position, chemical enrichment of the surface of B-type stars during the MS).

During the RSG phase, we use a mass-loss rate determined on the basis of observations by Crowther \cite*{Crowther2001a}, which is similar to the rate of de Jager et al. \cite*{deJager1988a}. However, for the most massive stars of this grid ($M_\text{ini} \geq 20\,M_\odot$), we observed that some layers in the outermost part of our models became significantly more luminous than the Eddington luminosity of the star, due to the strong opacity in the iron peak. In that case, this very unstable situation could favour mass loss. We thus decided to increase the standard mass-loss rate by a factor of $3$ in that case. Interestingly enough, it led to mass-loss rates comparable to the ones determined by van Loon et al. \cite*{vanLoon2005a}. This kind of increased mass-loss rates during the RSG phase was also used in previous works \cite{vanBeveren1998a,vanBeveren1998b}, and more recently in Chieffi \& Limongi \cite*{Chieffi2013a}, that led to similar conclusions.

\subsubsection{Consequences on RSGs}

\begin{figure}
\begin{center}
\includegraphics[width=.75\textwidth]{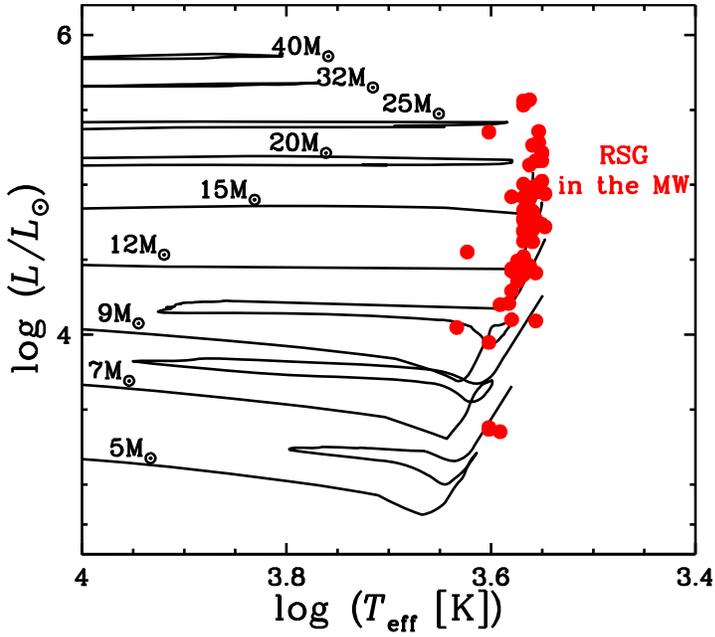}
\caption{Hertzsprung-Russell diagram tracks for rotating models ($Z_\odot$) between $5$ and $40\, M_\odot$ (Ekstr\"om et al., 2012). The red dots correspond to the position of Galactic RSGs (Levesque et al., 2005)}
\label{RSGPos}
\end{center}
\end{figure}

The inclusion of an increased mass-loss rate for the more massive RSGs at solar metallicity has as the direct consequence to help these models to evolve bluewards at some points during their RSG phase. On Fig.~\ref{RSGPos} are shown the tracks in the Hertzsprung-Russell diagram (HRD) for our rotating models between $5$ and $40\, M_\odot$. The red dots correspond to the positions of observed Galactic RSGs \cite{Levesque2005a}. First, we see that our models reproduce well the observations, particularly the upper luminosity of the RSGs.

The second interesting point is that rotating models of $20$ and $25\, M_\odot$ evolve away from the RSG branch at the end of their life. This is a direct consequence of the strong mass loss they encounter during the RSG stage.

\subsubsection{Blue-supergiants and Wolf-Rayet stars, and consequences on SN progenitors}

From our grid of rotating models, the following evolutionary scenarios are possible \cite{Georgy2012b}:
\begin{itemize}
\item MS $\rightarrow$ RSG for star with initial mass $M \lesssim 16.8\, M_\odot$ 
\item MS $\rightarrow$ RSG $\rightarrow$ nitrogen rich (WN) Wolf-Rayet star (WR) for star with initial mass $16.8\, M_\odot \lesssim M \lesssim 25.0 \, M_\odot$
\item MS $\rightarrow$ Yellow supergiant (YSG) $\rightarrow$ WN $\rightarrow$ carbon rich (WC) WR star for star with initial mass $25.8\, M_\odot \lesssim M \lesssim 60.0 \, M_\odot$
\item MS $\rightarrow$ WN $\rightarrow$ WC for star with initial mass $M \gtrsim 60.0 \, M_\odot$
\end{itemize}

Interestingly, the models between $20$ and $25\, M_\odot$ end their life as WN stars (or even as luminous blue variable (LBV) stars, see Groh et al. \cite*{Groh2013a}), with $\log(T_\text{eff})\sim 4.1 - 4.3$. This is in good agreement with the identified progenitor of SN 2008ax \cite{Crockett2008a}. The estimated maximal mass we obtain for a type IIP progenitor of $16.8\, M_\odot$ \cite{Georgy2012b} is also in good agreement with the work of Smartt et al. \cite*{Smartt2009a}, who determine a maximal initial mass of $16.5 \pm 1.5\, M_\odot$ for the progenitors of this kind of Supernovae (SNe) (Note however that the so-called ``red-supergiant problem'' could have other explanations, see Walmswell \& Eldridge \cite*{Walmswell2012a}).

\subsection{Strong mass-loss rates for less massive RSGs?}

\begin{figure}
\begin{center}
\includegraphics[width=.75\textwidth]{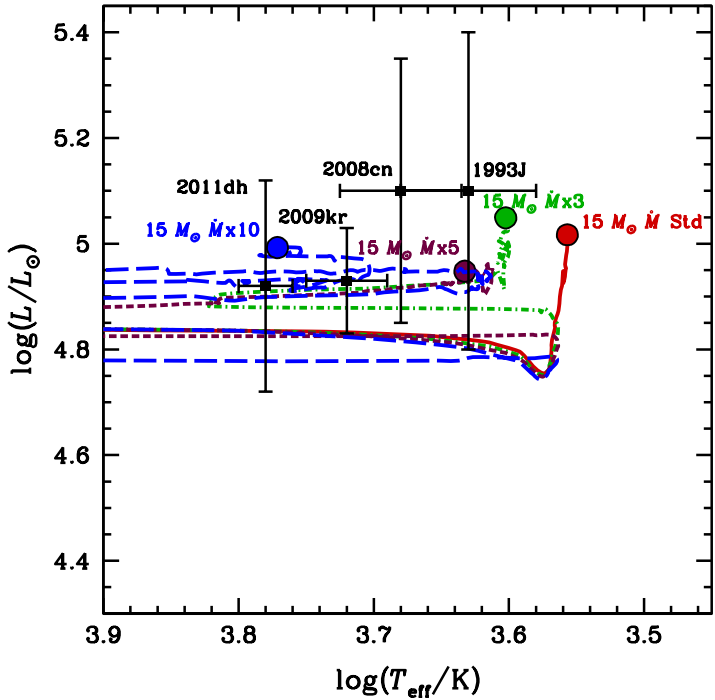}
\caption{HRD tracks for $15\, M_\odot$ rotating models ($Z_\odot$, Georgy, 2012) with various mass-loss rates during the RSG phase: standard one (red solid curve), $3$ times (green dot-dashed curve), $5$ times (purple short-dashed curve) and $10$ times standard one (blue long-dashed curve). The observational data are from Maund et al. (2004, 2011), and Elias-Rosa et al. (2009, 2010).}
\label{YSG}
\end{center}
\end{figure}

During the past 20 years, 3 progenitor stars have been identified for some (rare) type IIL or IIb SNe: SN 1993J \cite{Maund2004a}, SN 2009kr \cite{Elias-Rosa2010a}, and SN 2011dh \cite{Maund2011a}, as well as a peculiar progenitor for the type IIP SN 2008cn \cite{Elias-Rosa2009a}. All these progenitors are YSGs, with $3.63 \leq \log(T_\text{eff}) \leq 3.78$, and a luminosity of $\log(L/L_\odot) \sim 5$. The luminosity indicates a star with an initial mass of $\sim 15\, M_\odot$. However, single star models show that with such a mass, the star ends its life as a type IIP SN. Binarity could play an important role in the evolution of such stars, as the progenitor of SN 1993J is most probably a binary star \cite{Podsiadlowski1993a}, and this possibility is also not excluded for SNe 2008cn and 2009kr. However, the observations for SN 2011dh do not indicate any hint of binarity \cite{Maund2011a}. A recent work by Benvenuto et al. \cite*{Benvenuto2013a} shows that the secondary star could be hidden, thus the absence of a companion cannot be taken as a proof that the star did not interact with a companion in the past.

We can however wonder if a single star model, with an increased mass-loss rate during the RSG stage, is also able to reproduce the position and characteristics of these progenitors (see \cite{Bersten2012a}). In Georgy \cite*{Georgy2012a}, we have shown that it is indeed the case. Fig~\ref{YSG} shows the tracks in the HRD for a $15\, M_\odot$ model at solar metallicity, with a standard mass-loss rate (red solid track), or with a mass-loss rate increased by a factor of $3$ (green dot-dashed track), $5$ (purple short-dashed track) or $10$ (blue long-dashed track), during the RSG phase. We see that the effect of increasing the mass-loss rate is to shift the final position of the model towards higher and higher $T_\text{eff}$. The final properties of the progenitor star (core mass, ejected H mass) are also compatible with the values deduced from the SN light curve \cite{Bersten2012a}.

Note also here that increasing the mass-loss rate during the RSG phase shortens this phase, and thus the duration of the episode with strong mass loss. These effects compensate somehow, and  the models with $3$ to $10$ times the standard $\dot{M}$ finish with roughly the same final mass. Our computations suggest that this effect forbids a later evolution to a WC star for these lower-mass RSGs (see Meynet et al. in these proceedings). This last point is however sensitive to the detailed mass-loss history, as some other authors \cite{vanBeveren1998a,vanBeveren1998b} obtained such WR stars for models with initial mass as low as $15\, M_\odot$.

\subsection{How to distinguish a blue star coming from the red and an``ever-blue'' star?}

Is there a way to determine whether a blue star evolved from a previous RSG phase? For example, is it possible to determine observational properties that are unique to a star coming from the red part of the HRD compared to another one, which has still not crossed it?

A first point to notice is that the surface chemical composition would be different. A star which has been previously in a RSG phase has encountered the first dredge up due to the large convective zone developing during this phase. The effect is to bring to the stellar surface products of H-burning through CNO cycle, mainly, a strong enrichment of He and N, and a depletion in H, C and O. Table~\ref{Compo} shows the abundances of these elements, as well as the ratio $^{12}$C/$^{13}$C for the non-rotating  and rotating $25\, M_\odot$ models of Ekstr\"om et al. \cite*{Ekstrom2012a}.

\begin{figure}
\begin{center}
\includegraphics[width=.75\textwidth]{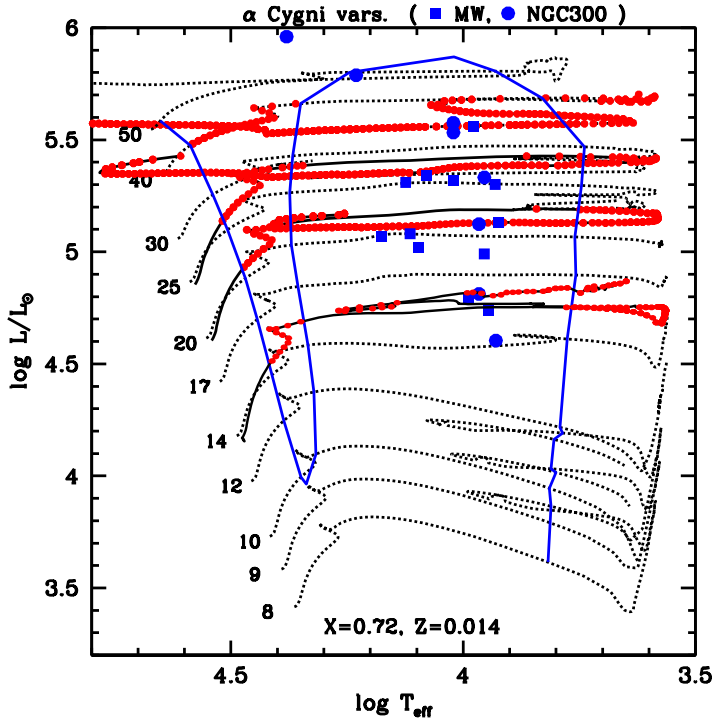}
\caption{HRD tracks for various models (non-rotating ones in dotted lines, rotating ones in solid lines). The $14\, M_\odot$ model was computed with a mass-loss rate increased by a factor of $5$ during the RSG phase. The red dots indicate models for which at least one radial pulsation mode is excited (Saio et al., submitted). The blue points are observed $\alpha$ Cyg variables in the Galaxy (Kudritzki et al., 1999; Przybilla et al., 2010; Firnstein and Przybilla, 2012) and in NGC 300 (Bresolin et al., 2004; Kudritzki et al., 2008). Blue line indicates $\beta$ Cephei instability region ($\log(T_\text{eff}) \sim 4.3-4.5)$  and the blue-edge of the Cepheid instability strip ($\log(T_\text{eff}) \sim 3.8$).}
\label{Pulsation}
\end{center}
\end{figure}

Another indication could come from the pulsational properties of the star. On Fig.~\ref{Pulsation}, we indicate by red dots models which have at least one radial pulsational mode, for four different initial masses. An interesting thing is that during the first crossing of the HRD, no modes are excited, and the star should not vary. However, several $\alpha$ Cyg variables are found in the corresponding $T_\text{eff}$ region. On the contrary, if the star crosses again the HRD towards the blue side, the higher $M/L$ ratio (due to the strong mass loss during the RSG phase) allows some pulsational modes to be excited in the $T_\text{eff}$ range between $\sim 3.7$ and $\sim 4.3$. It could thus be possible to discriminate between these two types of stars on the basis of their pulsational properties. This will be studied in detail in a future work (Saio et al., submitted).

Finally, due to the strong breaking encountered during the RSG phase, stars coming back from the red are probably slow rotators, contrarily to the stars which have still not crossed the HRD, which can have any rotational velocities.

\begin{table}
\begin{center}
\caption{Main abundances before and after the RSG phase of a non-rotating and a rotating $25\, M_\odot$ model. By ``before'' and ``after'', we mean the first and last time that $\log(T_\text{eff}) < 3.66$ (roughly corresponding to the temperature at which the surface convective zone begins to sink).}
\label{Compo}
\begin{tabular}{c|cc||cc}
& \multicolumn{2}{c}{non rotating} & \multicolumn{2}{c}{rotating}\\
element & before RSG & after RSG  & before RSG & after RSG\\
\hline
\rule{0pt}{11pt}$^1$H & $0.720$ & $0.450$ & $0.638$ & $0.370$\\
$^4$He & $0.266$ & $0.536$ & $0.348$ & $0.616$\\
$^{12}$C & $2.28\cdot 10^{-3}$ & $5.52\cdot 10^{-5}$ & $1.07\cdot 10^{-3}$ & $1.97\cdot 10^{-4}$\\
$^{14}$N & $6.59\cdot 10^{-4}$ & $7.32\cdot 10^{-3}$ & $3.54\cdot 10^{-3}$ & $6.51\cdot 10^{-3}$\\
$^{16}$O & $5.72\cdot 10^{-3}$ & $1.10\cdot 10^{-3}$ & $3.94\cdot 10^{-3}$ & $1.81\cdot 10^{-3}$\\
$^{12}$C/$^{13}$C & $81.43$ & $3.58$ & $9.47$ & $5.42$\\
\end{tabular}
\end{center}
\end{table}

\section{Conclusions}

The mass lost during the RSG phase is a key parameter determining the later evolution of massive stars going through such a phase. We have seen that increasing the standard mass-loss rates during the RSG phase for the most massive RSGs allows for a good fit of the maximal luminosity of Galactic RSGs, and also of the maximal initial mass leading to a type IIP SN. At lower mass, it also provides an alternative scenario to the binary channel to explain the peculiar position in the HRD of the progenitors of type IIL or IIb SNe.

Finally, we also propose some observational tests which could make it possible to distinguish the blue stars that evolve from a previous RSG phase and other ones, that did not have a previous RSG stage.

\bibliographystyle{astron}
\bibliography{MyBiblio}

\end{document}